\documentstyle[12pt]{article}
\def\PRL #1 #2 #3{{\sl Phys. Rev. Lett.} {\bf#1} (#2) #3}
\def\NPB #1 #2 #3{{\sl Nucl. Phys.} {\bf B#1} (#2) #3}
\def\NPBFS #1 #2 #3 #4{{\sl Nucl. Phys.} {\bf B#2} [FS#1] (#3) #4}
\def\CMP #1 #2 #3{{\sl Commun. Math. Phys.} {\bf #1} (#2) #3}
\def\PRD #1 #2 #3{{\sl Phys. Rev.} {\bf D#1} (#2) #3}
\def\PLA #1 #2 #3{{\sl Phys. Lett.} {\bf #1A} (#2) #3}
\def\PLB #1 #2 #3{{\sl Phys. Lett.} {\bf #1B} (#2) #3}
\def\JMP #1 #2 #3{{\sl J. Math. Phys.} {\bf #1} (#2) #3}
\def\PTP #1 #2 #3{{\sl Prog. Theor. Phys.} {\bf #1} (#2) #3}
\def\SPTP #1 #2 #3{{\sl Suppl. Prog. Theor. Phys.} {\bf #1} (#2) #3}
\def\AoP #1 #2 #3{{\sl Ann. of Phys.} {\bf #1} (#2) #3}
\def\PNAS #1 #2 #3{{\sl Proc. Natl. Acad. Sci. USA} {\bf #1} (#2) #3}
\def\RMP #1 #2 #3{{\sl Rev. Mod. Phys.} {\bf #1} (#2) #3}
\def\PR #1 #2 #3{{\sl Phys. Reports} {\bf #1} (#2) #3}
\def\AoM #1 #2 #3{{\sl Ann. of Math.} {\bf #1} (#2) #3}
\def\UMN #1 #2 #3{{\sl Usp. Mat. Nauk} {\bf #1} (#2) #3}
\def\FAP #1 #2 #3{{\sl Funkt. Anal. Prilozheniya} {\bf #1} (#2) #3}
\def\FAaIA #1 #2 #3{{\sl Functional Analysis and Its Application} {\bf
#1} (#2) #3}
\def\BAMS #1 #2 #3{{\sl Bull. Am. Math. Soc.} {\bf #1} (#2)
#3} \def\TAMS #1 #2 #3{{\sl Trans. Am. Math. Soc.} {\bf #1} (#2) #3}
\def\InvM #1 #2 #3{{\sl Invent. Math.} {\bf #1} (#2) #3}
\def\LMP #1 #2 #3{{\sl Letters in Math. Phys.} {\bf #1} (#2) #3}
\def\IJMPA #1 #2 #3{{\sl Int. J. Mod. Phys.} {\bf A#1} (#2) #3}
\def\AdM #1 #2 #3{{\sl Advances in Math.} {\bf #1} (#2) #3}
\def\RMaP #1 #2 #3{{\sl Reports on Math. Phys.} {\bf #1} (#2) #3}
\def\IJM #1 #2 #3{{\sl Ill. J. Math.} {\bf #1} (#2) #3}
\def\APP #1 #2 #3{{\sl Acta Phys. Polon.} {\bf #1} (#2) #3}
\def\TMP #1 #2 #3{{\sl Theor. Mat. Phys.} {\bf #1} (#2) #3}
\def\JPA #1 #2 #3{{\sl J. Physics} {\bf A#1} (#2) #3}
\def\JSM #1 #2 #3{{\sl J. Soviet Math.} {\bf #1} (#2) #3}
\def\MPLA #1 #2 #3{{\sl Mod. Phys. Lett.} {\bf A#1} (#2) #3}
\def\JETP #1 #2 #3{{\sl Sov. Phys. JETP} {\bf #1} (#2) #3}
\def\JETPL #1 #2 #3{{\sl  Sov. Phys. JETP Lett.} {\bf #1} (#2) #3}
\def\PHSA #1 #2 #3{{\sl Physica} {\bf A#1} (#2) #3}
\def\CQG #1 #2 #3{{\sl Class. Quantum Grav.} {\bf #1} (#2) #3}
\def\SJNP #1 #2 #3{{\sl Sov. J. Nucl. Phys. (Yadern.Fiz.)} {\bf #1} (#2) #3}
\def\a{\alpha}\def\b{\beta}\def\d{\delta}

\def\Th{\Theta}

\newcommand{\p}[1]{(\ref{#1})}
\begin{document}
\thispagestyle{empty}
\renewcommand{\thefootnote}{\fnsymbol{footnote}}
\begin{flushright}
Preprint DFPD 97/TH/05\\
hep-th/9701149\\
January, 1997
\end{flushright}

\vspace{3truecm}
\begin{center}
{\large\bf Covariant Action for the Super--Five--Brane
of M--Theory}

\vspace{1cm}
Igor Bandos$^1$\footnote{e--mail: kfti@rocket.kharkov.ua}, Kurt
Lechner$^2$\footnote{e--mail: lechner@pd.infn.it}, Aleksei
Nurmagambetov$^{1*}$\\
Paolo
Pasti$^2$\footnote{e--mail:  pasti@pd.infn.it}, Dmitri
Sorokin$^{1*}$ and Mario Tonin$^2$ \footnote{e--mail:
tonin@pd.infn.it}

\vspace{0.5cm}
$^1${\it National Science Center\\
Kharkov Institute of Physics and Technology,\\
Kharkov, 310108, Ukraine}

\bigskip
$^2${\it Universit\`a Degli Studi Di Padova,
Dipartimento Di Fisica ``Galileo Galilei''\\
ed INFN, Sezione Di Padova,
Via F. Marzolo, 8, 35131 Padova, Italia}

\vspace{1.cm}
{\bf Abstract}
\end{center}
We propose a complete, d=6 covariant and kappa--symmetric, action
for an M--theory five--brane propagating
in $D=11$ supergravity background.



\renewcommand{\thefootnote}{\arabic{footnote}}
\setcounter{footnote}0
\newpage
Among new types of super--p--branes \cite{le}--\cite{hs}, that have
attracted a lot of attention during last several years, a five--brane
\cite{ds,gt} of eleven--dimensional M-theory \cite{M} is one of few for
which the complete $\kappa$--invariant action has been unknown. In
\cite{pst} a covariant action for the bosonic five--brane interacting with
gravitational and antisymmetric fields of $D=11$ supergravity was
constructed, which completed partial results on the structure of the
bosonic part of the action for the five--brane of M-theory
\cite{5}--\cite{ps}.

In the present letter we propose a manifestly covariant
$\kappa$--symmetric action for a five--brane propagating in $D=11$
superspace of M--theory.  Since the five--brane carries in its worldvolume
a self--dual rank--two field, to construct the action we use a
Lorentz--covariant approach to describing self--dual gauge fields proposed
and developed in \cite{pst0}\footnote{When this paper was prepared for
publication we learned that in a noncovariant formulation \cite{ps} the
proof of the $\kappa$-invariance of a super--five--brane action was also
carried out by J. H. Schwarz with collaborators \cite{s}.}.  Only results
are presented, while a detailed proof of $\kappa$--invariance is postponed
to a forthcoming paper.

The five--brane action has the same form as in the bosonic case \cite{pst}
but with $D=11$ background fields replaced with superfields in
curved $D=11$ superspace parametrized by bosonic coordinates
$X^{\underline m}$ (${\underline m}=0,1...,10$) and Grassmann spinor
coordinates $\Th^{\underline\mu}$ (${\underline\mu}$=1,...,32) called
altogether as $Z^{\underline M}$ \footnote{We use underlined indices for
denoting the coordinates of target superspace and not underlined ones for
the coordinates of the five--brane worldvolume. The signature of the
metrics is chosen almost negative, the external derivative acts from
the right and the $D=11$ gamma--matrices are imaginary}:
$$ S=-\int d^6x\left[ \sqrt{-det(g_{mn}+i\tilde
H_{mn})}-\sqrt{-g}{1\over {4\partial_r a\partial^r
a}}\partial_la(x)H^{*lmn}H_{mnp}\partial^pa(x)\right] $$
\begin{equation}\label{ac}
-\int\left[C^{(6)}+{1\over 2}F\wedge C^{(3)}\right].
\end{equation}
In \p{ac} small $x^m$ ($m$=0,1,...,5) parametrize five--brane
worldvolume; $a{(x)}$ is an auxiliary scalar field which ensures $d=6$
general coordinate invariance of the action \cite{pst};
\begin{equation}\label{met}
g_{mn}(x)=E_m^{{\underline
a}}(x)\eta_{{\underline{ab}}}E_n^{\underline a}(x)\qquad
({\underline a},{\underline b}=0,...,10)
\end{equation}
is an induced worldvolume metric constructed of components of
the $D=11$ supervielbeins $E^{\underline A}=dZ^{\underline M}
E_{\underline M}^{~\underline A}$ pulled back to the worldvolume
($\underline A=(\underline a, \underline\a)$ denote tangent superspace
indices.). In the flat target superspace the metric takes the form
\begin{equation}\label{metf}
g_{mn}(x)=\partial_m\Pi^{{\underline
m}}(x)g_{{\underline{mn}}}\partial_n\Pi^{\underline n}(x)\qquad
(\Pi^{\underline m}=dX^{\underline m}(x)-id\Theta\Gamma^{\underline
m}\Theta).
\end{equation}
$A_{mn}(x)$ is a
worldvolume self--dual (or so--called chiral)
field with the field strength
$F_{mnl}=2(\partial_{l}A_{mn}+ \partial_{m}A_{nl}+\partial_{n}A_{lm}),$
(note that a generalized self--duality condition for $A_{mn}$ arises from
\p{ac} as an equation of motion \cite{ps,pst});
\begin{equation}\label{h} H_{lmn}(x)=F_{lmn}-C^{(3)}_{lmn}, \qquad
\tilde
H_{mn}\equiv{1\over{\sqrt{-(\partial a)^2}}}H^{*}_{mnl}\partial^la(x),
\qquad H^{*mnl}={1\over{3!\sqrt{-g}}}\varepsilon^{mnlpqr}H_{pqr},
\end{equation}
and $C^{(3)}_{lmn}$ and $C^{(6)}_{lmnpqr}$ are pullbacks into worldvolume
of superforms  $C^{(3)}(X,\Theta)$ and $C^{(6)}(X,\Theta)$ of $D=11$
supergravity whose field strengths are dual to each other in the following
sense \cite{cl}
\begin{equation}\label{c}
~^*dC^{(3)}=dC^{(6)}+{1\over 2}C^{(3)}R^{(4)}
\equiv R^{(7)}, \qquad R^{(4)}\equiv dC^{(3)},
\end{equation}
(where $^*$ denotes eleven--dimensional "bosonic" Hodge operation
accompanied by \linebreak $
(\Gamma_{\underline{a}\underline{b}})_{\underline{\alpha}
\underline{\beta}}
~~\rightarrow~~
(\Gamma_{\underline{a_1}...\underline{a_5}})_{\underline{\alpha}
\underline{\beta}}$).
The $D=11$ supergravity background fields are
assumed to satisfy the constraints:
$$ T^{\underline a}={\cal
D}E^{\underline a}=-iE^{\underline \alpha}\wedge E^{\underline
\beta}\Gamma_{\underline{\alpha}\underline{\beta}}^{\underline a}+
E^{\underline b}\wedge E^{\underline \beta}
T_{\underline{b\beta}}^{\underline a}
+{1\over 2} E^{\underline b}\wedge E^{\underline c}
T_{\underline{b}\underline{c}}^{\underline a},
$$
$$
R^{(4)}=dC^{(3)}={1\over 2}E^{\underline b}\wedge E^{\underline a}\wedge
E^{\underline\alpha}\wedge E^{\underline \beta}
(\Gamma_{\underline{ab}})_{\underline{\alpha\beta}}
$$
\begin{equation}\label{con}
+{1\over {4!}}E^{\underline a}\wedge E^{\underline b}\wedge
E^{\underline c}\wedge E^{\underline d}R_{\underline{d}\underline{c}
\underline{b}\underline{a}},
\end{equation}
$$
R^{(7)}={i\over {5!}}E^{\underline{a_1}}\wedge...\wedge
E^{\underline{a_5}}\wedge E^{\underline\alpha}\wedge E^{\underline \beta}
(\Gamma_{\underline{a_1}...\underline{a_5}})_{\underline{\alpha}
\underline{\beta}}+{\cal O}\left((E^{\underline{a}})^6\right).
$$

Local transformations which leave the action \p{ac} invariant were
discussed in \cite{pst}, so we only present a local symmetry which
reflects an auxiliary role of $a(x)$ $$ \delta a(x)=\varphi(x), $$
\begin{equation}\label{phi}
\delta A_{mn}={{\varphi(x)}\over {2(\partial
a)^2}}(H_{mnp}\partial^pa-{\cal V}_{mn}),
\end{equation} where
$$
{\cal
V}^{mn}\equiv-2\sqrt{{{(\partial a)^2}\over{g}}}
{{\delta{\sqrt{-\det(g_{pq}+i\tilde H_{pq})}}}\over{\delta\tilde H_{mn}}}.
$$

The second integral in \p{ac} is the Wess--Zumino term
$\int{\cal L}^{(6)}_{WZ}$. Its external
derivative (required for proving the $\kappa$--invariance) is a closed
7-superform
\begin{equation}\label{wz} d{\cal
L}^{(6)}_{WZ}=R^{(7)}+{1\over{2}}H\wedge R^{(4)},
\end{equation} which in flat
$D=11$ superspace takes the following form:
\begin{equation}\label{wzf}
d{\cal L}^{(6)}_{WZ}={i\over {5!}}\Pi^{{\underline m}_1}\wedge...\wedge
\Pi^{{\underline m}_5}d\Theta\Gamma_{{\underline m}_1...{\underline
m}_5}d\Theta+{1\over{2}}H\wedge\Pi^{{\underline m}_1}\wedge
\Pi^{{\underline m}_2}d\Theta\Gamma_{{\underline m}_1{\underline
m}_2}d\Theta.
\end{equation}
Note that the coefficient in front of the Wess--Zumino term is fixed
already in the purely bosonic case by the requirement of the invariance of
\p{ac} under \p{phi} (see \cite{pst}).

As in the case of the D-branes \cite{c} an indication that the action
\p{ac} is invariant under fermionic $\kappa$--transformations is the
existence of a matrix $\bar\Gamma$ whose square is the unit matrix.
In our case a relevant matrix has the following form:
$$
\sqrt{-\det(g+i\tilde H)}\bar\Gamma={\sqrt{-g}}\big[
\Gamma^{(6)}+{i\over{2\sqrt{-(\partial a)^2}}}
{\tilde
H}^{mn}\Gamma_{mn}\Gamma_{p}\partial^pa
$$
\begin{equation}\label{G}
+{1\over
{8^(\partial a)^2}}\partial^{m_1}a
\varepsilon_{m_1...m_6}{\tilde H}^{m_2m_3}{\tilde
H}^{m_4m_5}\Gamma^{m_6}\Gamma_p\partial^pa\big],
\end{equation}
where
$$
\Gamma_m=\Gamma_{\underline a}E^{\underline a}_m \qquad
(\Gamma_m=\Gamma_{\underline n}\Pi^{\underline n}_m~~{\rm in~flat~target
~superspace})
$$
are the pullbacks into the worldvolume of the $D=11$ gamma--matrices,
$\Gamma^{(n)}$ is the antisymmetriezed product of $n$ $\Gamma_m$.

The action \p{ac} indeed possesses $\kappa$--invariance, the
$\kappa$--transformations of the worldvolume fields being:
$$
i_{\kappa}E^{\underline\a}=\delta_\kappa Z^ME_M^{\underline\alpha}
=\kappa^{\underline\b}(1+\bar\Gamma)^{~\underline\b}_{\underline\a},
\qquad
i_{\kappa}E^{\underline a}=0, \qquad
\d_{\kappa}g_{mn}=-4iE_{\{m}\Gamma_{n\}}i_{\kappa}E,
$$
\begin{equation}\label{k}
\d_{\kappa}H=-i_{\kappa}dC^{(3)},
\end{equation}
or in the flat case
$$
\d_{\kappa}\Theta^{\underline\mu}
=\kappa^{\underline\nu}(1+\bar\Gamma)^{~\underline\mu}_{\underline\nu},
\qquad
\d_{\kappa}\Pi^{\underline m}=-2id\Theta\Gamma^{\underline
m}\d_{\kappa}\Theta, \qquad
\d_{\kappa}g_{mn}=-4i\partial_{\{m}\Theta\Gamma_{n\}}\d_{\kappa}\Theta,
$$
\begin{equation}\label{kf}
\d_{\kappa}H=-\Pi^{\underline n}\wedge\Pi^{\underline m}\wedge
d\Theta\Gamma_{{\underline{mn}}}\delta_{\kappa}\Theta.
\end{equation}

Because of a Born--Infeld--like form of \p{ac}
the check of the $\kappa$--invariance of the five--brane action
is carried out using the way analogous to that for the Dirichlet branes
\cite{c}. A difference is in the presence in the first integral of
\p{ac} of the term quadratic in $H$ whose $\kappa$--variation contributes
to the variation of the Wess--Zumino term.
As in the bosonic case \cite{pst}, upon a double dimensional reduction the
D=11 super--five--brane should reduce to a dual version of a D=10
Dirichlet super--4--brane. A detailed analysis of the action \p{ac} will
be made in a forthcoming paper.

In conclusion we have constructed the covariant action for the
five--brane of M--theory which is invariant under the $\kappa$--symmetry
transformations and contains the auxiliary scalar field $a(x)$.
The role of the auxiliary field is to ensure the covariance of the model
under $d=6$ worldvolume diffeomorphisms
(which makes the analysis of the model
much simpler) \footnote{In this sense $a(x)$ is analogous to auxiliary
(or compensator) fields of supergravity
models, whose presence enables one to make supersymmetry manifest.}, its
variation does not lead to independent field equations, and
$\partial_ma(x)$ cannot square to zero \cite{pst0,pst}. Thus the presence
of this field in the action might be a manifestation of nontrivial
topological features of the five--brane and M--theory itself.

\bigskip
\noindent
{\bf Acknowledgements}. Authors are grateful to Alvaro Restuccia, Mees de
Roo and Kellogg Stelle for discussion.  Work of K.L., P.P. and M.T. was
supported by the European Commission TMR programme ERBFMRX--CT96--045 to
which K.L., P.P. and M.T. are associated. I.B., A.N. and D.S. acknowledge
partial support from the grant N2.3/644
of the Ministry of Science and Technology of Ukraine and
the INTAS Grants N 93--127, N 93--493, and N 94--2317.

\end{document}